\documentclass[11pt,titlepage,a4paper]{article}
\usepackage{bozhomac}	
\usepackage{bozhlogo}	
\usepackage{cite}	

\title{\bfseries	\vspace*{-1.678902345in}
{\huge Heisenberg relations in the general case}
}

\vspace{1.7ex}

\author{
Bozhidar Z.\ Iliev
\thanks{Laboratory of Mathematical Modeling in Physics,
Institute for Nuclear Research and \mbox{Nuclear} Energy,
Bulgarian Academy of Sciences,
Boul.\ Tzarigradsko chauss\'ee~72, 1784 Sofia, Bulgaria}
\thanks{E-mail address: bozho@inrne.bas.bg}
\thanks{URL: http://theo.inrne.bas.bg/$\sim$bozho/}
}

\date{
 \vspace{2.27ex}\ShortTitle{Heisenberg relations}	\\[0.27ex]
 \vspace{3.27ex}
\small
	\begin{tabular}{r@{$\colon\to~$}l}
\vspace{0.09ex} Basic ideas	& 1997-2008	\\[0.09ex]
 \vspace{0.09ex} Began		& August 8, 2008\\[0.09ex]
 \vspace{0.09ex} Ended		& August 13, 2008\\[0.09ex]
 \vspace{0.27ex} Produced	& \fbox{\today}	\\[0.27ex]
	\end{tabular} \\[1.27ex]
\normalsize
\vspace{0.27ex}
\textsl{\bfseries
Report presented at the 9$^\text{the}$ International Workshop on\\
``Complex Structures, Integrability and Vector Fields''\\
Sofia, Bulgaria, 25 -- 29 August, 2008}		\\[2ex] %
 \small
	\begin{tabular}{r@{$\colon~$}l}
\normalsize\sffamily\bfseries
 \vspace{0.27ex} http://arXiv.org e-Print archive No. &
\normalsize\sffamily\bfseries
0809.0174 [math-ph] \\[1.27ex]
	\end{tabular} \\[-0.27ex]
\normalsize
 \vspace{4.27ex}{\Huge\BOZHO}	\\[4.27ex]
	\begin{tabular}{r@{\hspace{0.512em}}|@{\hspace{0.512em}}l}
\vspace{0.27ex}	
&
\vspace{0.27ex}
	\end{tabular} \\[1.27ex]
\vspace{0.27ex}	\\[0.27ex]
}

	\newcommand{\base}{\mathit{M}}	

\begin{document}		

\renewcommand{\thepage}{\roman{page}}

\renewcommand{\thefootnote}{\fnsymbol{footnote}} 
\maketitle				
\renewcommand{\thefootnote}{\arabic{footnote}}   

\tableofcontents		


\begin{abstract}

	The Heisenberg relations are derived in a quite general setting when
the field transformations are induced by three representations of a given group.
They are considered also in the fibre bundle approach. The results are
illustrated in a case of transformations induced by the Poincar\'e group.

\end{abstract}

\renewcommand{\thepage}{\arabic{page}}


\section {Introduction}
\label{Introduction}

	As Heisenberg relations or equations in quantum field theory are known a
kind of commutation relations between the field operators and the generators (of
a representation) of a group acting on system's Hilbert space of states. Their
(global) origin is in equations like
    \begin{equation}    \label{2.8-2}
\varphi'_i(r) = U\circ \varphi_i(r)\circ U^{-1} .
    \end{equation}
which connect the components $\varphi_i$ and $\varphi'_i$ of a quantum field
$\varphi$ with respect to two frames of reference. Here $U$ is an operator
acting on the state vectors of the quantum system considered and it is
expected that the transformed field operators  $\varphi'_i$ can be expressed
explicitly by means of $\varphi_i$ via some equations. If the
elements $U$ (of the representation) of the group are labeled by
$b=(b^1,\dots,b^s)\in\field^s$ for some $s\in\field[N]$ (we are dealing, in
fact, with a Lie group), \ie we may write $U(b)$ for $U$, then the
corresponding Heisenberg relations are obtained from~\eref{2.8-2} with $U(b)$
for $U$ by differentiating it with respect to $b^\omega$, $\omega=1,\dots,s$,
and then setting $b=b_0$, where $b_0\in\field^s$ is such that $U(b_0)$ is the
identity element.

	The above shows that the Heisenberg relations are from pure
geometric-group-theoretical origin and the only physics in them is the
motivation leading to equations like~\eref{2.8-2}. However, there are strong
evidences that to the Heisenberg relations can be given dynamical/physical
sense by identifying/replacing in them the generators (of the representation)
of the group by the corresponding operators of conserved physical quantities if
the system considered is invariant with respect to this group (see, \eg the
discussion in~\cite[\S~68]{Bjorken&Drell-2}).

	In sections~\ref{Subsect6.1}-\ref{Subsect6.3}, we consider
Heisenberg relations in the non-bundle approach. At first
(section~\ref{Subsect6.1}), we derive the Heisenberg relation connected with
the Poincar\'e group. Then (section~\ref{Subsect6.2}) the Heisenberg relations
arising from internal transformation, which are related with conserved charges,
are investigated. At last, in section~\ref{Subsect6.3} are considered the
Heisenberg relations in the most general case, when three representations of a
group are involved. In section~\ref{Subsect6.4} are investigated the
Heisenberg relation on the ground of fibre bundles. Section~\ref{Conclusion}
closes the paper.

\section{The Poincar\'e group}
\label{Subsect6.1}

	Suppose we study a quantum field with components $\varphi_i$ relative
to two reference frames connected by a general Poincar\'e
transformation
    \begin{equation}    \label{2.3-2}
u '(x)= \Lambda u(x)+a .
    \end{equation}
Here $x$ is a point in the Minkowski spacetime $M$, $u $ and $u'$ are the
coordinate homeomorphisms of some local charts in $M$, $\Lambda$ is a Lorentz
transformation (\ie a matrix of a 4\ndash rotation), and $a\in\field[R]^4$ is
fixed and represents the components of a 4\ndash vector translation. The
``global' version of the Heisenberg relations is expressed by the equation
	\begin{equation}	\label{6.1}
U(\Lambda,a)\circ\varphi_i(x)\circ U^{-1}(\Lambda,a)
=
{D}_i^j(\Lambda,a)\varphi_j(\Lambda x+a) ,
	\end{equation}
where $U$ (resp.\ $D$) is a representation of the Poincar\'e group on the space
of state vectors (resp.\ on the space of field operators),
 $U(\Lambda,a)$ (resp.\ $\Mat{D}(\Lambda,a)=[{D}_i^j(\Lambda,a)]$) is the
mapping (resp.\ the matrix of the mapping) corresponding via $U$ (resp.\ $D$)
to~\eref{2.3-2}. Note that here we have rigorously to write
$\varphi_{u,i}:=\varphi_i\circ u^{-1}$ for $\varphi_i$, \ie  we have omitted
the index $u$. Besides, the point $x\in\base$ is identified with
$\Mat{x}=u(x)\in\field[R]^4$. Since for $\Lambda=\openone$ and
$a=\Mat{0}\in\field[R]^4$ is fulfilled $u'(x)=u(x)$, we have
    \begin{equation}    \label{6.2}
U(\openone,\Mat{0}) = \id
\qquad
\Mat{D}(\openone,\Mat{0}) = \openone ,
    \end{equation}
where $\id$ is the corresponding identity mapping and $\openone$ stands for the
corresponding identity matrix. Let $\Lambda=[\Sprindex[\Lambda]{\nu}{\mu}]$,
 $\Lambda^{\mu\nu}:=\eta^{\nu\lambda}\Sprindex[\Lambda]{\lambda}{\mu}$, with
$\eta^{\mu\nu}$ being the components of the Lorentzian metric with signature
$(-+++)$, and define
    \begin{subequations}    \label{6.3}
    \begin{align}    \label{6.3a}
T_\mu
& := \frac{\pd U(\Lambda,a)}{\pd a^\mu}
\Big|_{(\Lambda,a)=(\openone,\Mat{0})}
\\		    \label{6.3b}
S_{\mu\nu}
& := \frac{\pd U(\Lambda,a)}{\pd \Lambda^{\mu\nu}}
\Big|_{(\Lambda,a)=(\openone,\Mat{0})}
\\		    \label{6.3c}
H_{j\mu}^{i}
& := \frac{\pd {D}_j^i(\Lambda,a)}{\pd a^\mu}
\Big|_{(\Lambda,a)=(\openone,\Mat{0})}
\\		    \label{6.3d}
I_{j\mu\nu}^{i}
& := \frac{\pd {D}_j^i(\Lambda,a)}{\pd \Lambda^{\mu\nu}}
\Big|_{(\Lambda,a)=(\openone,\Mat{0})} .
    \end{align}
    \end{subequations}
The particular form of the numbers $I_{j\mu\nu}^i$ depends on the field under
consideration. In particular, we have
    \begin{subequations}    \label{6.5}
    \begin{alignat}{2}
			    \label{6.5a}
& I_{1\mu\nu}^1 = 0
&\qquad &\text{for spin-0 (scalar) field}
\\			    \label{6.5b}
& I_{\rho\mu\nu}^\sigma
=
\delta_\mu^\sigma \eta_{\nu\rho} - \delta_\nu^\sigma \eta_{\mu\rho}
&\qquad &\text{for spin-1 (vector) field}
\\			    \label{6.5c}
& [ I_{j\mu\nu}^i ]_{i,j=1}^4
=
-\frac{1}{2} \iu \sigma_{\mu\nu}
&\qquad &\text{for spin-{\footnotesize $\frac{1}{2}$} (spinor) field} .
    \end{alignat}
    \end{subequations}

	Differentiating~\eref{6.1} relative to $a^\mu$ and setting after that
$(\Lambda,a)=(\openone,\Mat{0})$, we find
    \begin{equation}    \label{6.4}
[T_\mu,\varphi_i(x)]_{\_} = \pd_\mu\varphi_i(x) + H_{i\mu}^{j}\varphi_j(x) ,
    \end{equation}
where $[A,B]_{\_}:=AB-BA$ is the commutator of some operators or matrices $A$
and $B$.
	Since the field theories considered at the time being are invariant
relative to spacetime translation of the coordinates, \ie with respect to
$\Mat{x}\mapsto\Mat{x}+a$, further we shall suppose that
    \begin{equation}    \label{6.6}
H_{j\mu}^{i} = 0.
    \end{equation}
In this case equation~\eref{6.4} reduces to
    \begin{subequations}    \label{6.7}
    \begin{equation}    \label{6.7a}
[T_\mu,\varphi_i(x)]_{\_} = \pd_\mu\varphi_i(x).
    \end{equation}
Similarly, differentiation~\eref{6.1} with respect to
$\Lambda^{\mu\nu}$ and putting after that $(\Lambda,a)=(\openone,\Mat{0})$, we
obtain
    \begin{equation}    \label{6.7b}
[S_{\mu\nu},\varphi_i(x)]_{\_}
= x_\mu \pd_\nu\varphi_i(x) - x_\nu \pd_\mu\varphi_i(x) +
I_{i\mu\nu}^{j}\varphi_j(x)
    \end{equation}
    \end{subequations}
where $x_\mu:=\eta_{\mu\nu}x^\nu$. The equations~\eref{6.7} are
identical up to notation with~\cite[eqs.(11.70) and~(11.73)]{Bjorken&Drell-2}.
Note that for complete correctness one should write $\varphi_{u,i}(\Mat{x})$
instead of $\varphi_i(x)$ in~\eref{6.7}, but we do not do this to keep our
results near to the ones accepted in the physical
literature~\cite{Roman-QFT,Bjorken&Drell,Bogolyubov&Shirkov}.

	As we have mentioned earlier, the particular Heisenberg
relations~\eref{6.7} are from pure geometrical-group-theoretical origin. The
following heuristic remark can give a dynamical sense to them. Recalling that
the translation (resp.\ rotation) invariance of a (Lagrangian) field theory
results in the conservation of system's momentum (resp.\ angular momentum)
operator $P_\mu$ (resp.\ $M_{\mu\nu}$) and the correspondences
    \begin{equation}    \label{6.9}
\ih T_{\mu} \mapsto P_{\mu}
\qquad
\ih S_{\mu\nu} \mapsto M_{\mu\nu} ,
    \end{equation}
with $\hbar$ being the Planck's constant (divided by $2\pi$), one may suppose
the validity of the Heisenberg relations
    \begin{subequations}    \label{6.10}
    \begin{align}    \label{6.10a}
[P_\mu,\varphi_i(x)]_{\_}
&= \ih\pd_\mu\varphi_i(x)
\\		    \label{6.10b}
[M_{\mu\nu},\varphi_i(x)]_{\_}
&= \ih \{ x_\mu \pd_\nu\varphi_i(x) - x_\nu \pd_\mu\varphi_i(x) +
I_{i\mu\nu}^{j}\varphi_j(x) \} .
    \end{align}
    \end{subequations}
However, one should be careful when applying the last two equations  in the
Lagrangian formalism as they are external to it and need a particular proof in
this approach; \eg they hold in the free field
theory~\cite{Bjorken&Drell,bp-MP-book}, but a general proof seems to be missing.
In the axiomatic quantum field
theory~\cite{Roman-QFT,Bogolyubov&et_al.-AxQFT,Bogolyubov&et_al.-QFT} these
equations are identically valid as in it the generators of the translations
(rotations) are identified up to a constant factor with the components of the
(angular) momentum operator, $P_\mu=\ih T_\mu$ ($M_{\mu\nu}=\ih S_{\mu\nu}$).


\section{Internal transformations}
\label{Subsect6.2}

	In our context, an internal transformation is a change of the reference
frame $(u,\{e^i\})$, consisting of a local coordinate system $u$ and a frame
$\{e^i\}$ in some vector space $V$, such that the spacetime coordinates remain
unchanged. We suppose that $e^i \colon x\in M\mapsto e^i(x)\in V$, where $M$ is
the Minkowski spacetime and the quantum field $\varphi$ considered takes values
in $V$, \ie $\varphi \colon x\in M\mapsto \varphi(x)=\varphi_i(x)e^i(x)\in V$

	Let $G$ be a group whose elements $g_b$ are labeled by $b\in\field^s$
for some $s\in\field[N]$.~%
\footnote{~%
In fact, we are dealing with an $s$-dimensional Lie group and $b\in\field^s$ are
the (local) coordinates of $g_b$ in some chart on $G$ containing $g_b$ in its
domain.%
}
Consider two reference frames $(u,\{e^i\})$ and $(u',\{e^{\prime\,i}\})$, with
$u'=u$ and $\{e^i\}$ and $\{e^{\prime\,i}\}$ being connected via a matrix
$I^{-1}(b)$, where $I \colon G\mapsto \GL(\dim V,\field)$ is a matrix
representation of $G$ and $I\colon G\ni g_b\mapsto I(b)\in \GL(\dim V,\field)$.
The components of the fields, known as field operators, transform into
(cf.~\eref{2.8-2})
    \begin{equation}    \label{6.13}
\varphi'_{u,i}(r) = U(b)\circ\varphi_{u,i}(r)\circ U^{-1}(b)
    \end{equation}
where $U$ is a representation of $G$ on the Hilbert space of state vectors and
$U \colon G\ni g_b\mapsto U(b)$. Now the analogue of~\eref{6.1} reads
    \begin{equation}    \label{6.14}
U(b)\circ\varphi_{u,i}(r)\circ U^{-1}(b) = I_i^j(b) \varphi_{u,j} (r)
    \end{equation}
due to $u'=u$ in the case under consideration.

	Suppose $b_0\in\field^s$ is such that $g_{b_0}$ is the identity element
of $G$ and define
    \begin{equation}    \label{6.14-1}
Q_\omega : = \frac{\pd U(b)}{\pd b^\omega} \Big|_{b=b_0}
\qquad
I_{i \omega}^j : = \frac{\pd I_i^j(b)}{\pd b^\omega} \Big|_{b=b_0}
    \end{equation}
where $b=(b^1,\dots,b^s)$ and $\omega=1,\dots,s$. Then,
differentiation~\eref{6.14} with respect to $b^\omega$ and putting in the
result $b=b_0$, we get the following Heisenberg relation
    \begin{equation}    \label{6.15}
[ Q_\omega, \varphi_{u,i}(r) ]_{\_} = I_{i\omega}^j \varphi_{u,j}(r)
    \end{equation}
or, if we identify $x\in\base$ with $r=u(x)$ and omit the subscript $u,$
    \begin{equation}    \label{6.16}
[ Q_\omega, \varphi_{i}(x) ]_{\_} = I_{i\omega}^j \varphi_{j}(x) .
    \end{equation}

	To make the situation more familiar, consider the case of one\ndash
dimensional group $G$, $s=1$, when $\omega=1$ due to which we shall identify
$b^1$ with $b=(b^1)$. Besides, let us suppose that
    \begin{equation}    \label{6.17}
I(b) = \openone \exp(f(b)-f(b_0))
    \end{equation}
for some $C^1$ function $f$. Then~\eref{6.16} reduces to
    \begin{equation}    \label{6.18}
[ Q_1, \varphi_{i}(x) ]_{\_} = f'(b_0) \varphi_i(x) ,
    \end{equation}
where $f'(b):=\frac{\od f(b)}{\od b}$. In particular, if we are dealing with
phase transformations, \ie
    \begin{equation}    \label{6.19}
U(b) = \e^{\frac{1}{\iu e} b Q_1}
\quad
I(b) = \openone \e^{-\frac{q}{\iu e} b}
\qquad
b\in\field[R]
    \end{equation}
for some constants $q$ and $e$ (having a meaning of charge and unit charge,
respectively) and operator $Q_1$ on system's Hilbert space of states (having a
meaning of a charge operator), then~\eref{6.14} and~\eref{6.18} take the
familiar form~\cite[eqs.~(2.81) and~(2-80)]{Roman-QFT}
    \begin{align}    \label{6.20}
& \varphi'_i(x)
=
\e^{\frac{1}{\iu e} b Q_1}\circ \varphi_i(x) \circ \e^{-\frac{1}{\iu e} b Q_1}
=
\e^{-\frac{q}{\iu e} b} \varphi(x)
\\		    \label{6.21}
& [Q_1,\varphi_i(x)]_{\_} = -q \varphi_i(x) .
    \end{align}

	The considerations in the framework of Lagrangian formalism invariant
under  phase
transformations~\cite{Roman-QFT,Bjorken&Drell,Bogolyubov&Shirkov} implies
conservation of the charge operator $Q$ and suggests the correspondence
(cf.~\eref{6.9})
    \begin{equation}    \label{6.22}
Q_1 \mapsto Q
    \end{equation}
which in turn suggests the Heisenberg relation
    \begin{equation}    \label{6.23}
[Q,\varphi_i(x)]_{\_} = - q \varphi_i(x) .
    \end{equation}
We should note that this equation is external to the Lagrangian formalism and
requires a proof in it~\cite{bp-MP-book}.


\section{The general case}
\label{Subsect6.3}

	The corner stone of the (global) Heisenberg relations is the equation
    \begin{equation}    \label{6.24}
 U\circ\varphi_{u,i}(r)\circ U^{-1}
=
\frac{\pd( u '\circ u ^{-1})(r)}{\pd r}
	\bigl( A^{-1}( u ^{-1}(r)) \bigr)_i^j
	\varphi_{u,j}( ( u '\circ u ^{-1})(r) )
    \end{equation}
representing the components $\varphi'_{u',i}$ of a quantum field $\varphi$ in a
reference frame $(u,\{e^{\prime\, i}=A_j^i e^j\})$ via its components
$\varphi_{u,i}$ in a frame $(u,\{e^i\})$ in two different way. Here $A=[A_i^j]$
is a non\ndash degenerate matrix\ndash valued function, $r\in\field[R]^4$ and
$\varphi_{u,i}:=\varphi_i\circ u^{-1}$. Now, following the ideas at the
beginning of section~\ref{Introduction}, we shall demonstrate how from the last
relation can be derived Heisenberg relations in the general case.

	Let $G$ be an  $s$-dimensional, $s\in\field[N]$, Lie group. Without
going into details, we admit that its elements are labeled by
$b=(b^1,\dots,b^s)\in\field^s$ and $g_{b_0}$ is the identity element of $G$ for
some fixed $b_0\in\field^s$. Suppose that there are given three representations
$H$, $I$ and $U$ of $G$ and consider frames of reference with the following
properties:
    \begin{enumerate}
\item
\(
H\colon G\ni g_b\mapsto H_b
\colon \field[R]^{\dim \base}\to \field[R]^{\dim \base}
\) and any change $(U,u)\mapsto(U',u')$ of the charts of $M$ is such
that $u'\circ u^{-1}=H_b$ for some $b\in\field^s$.

\item
$I \colon G\ni g_b\mapsto I(b)\in\GL(\dim V,\field)$
and any change $\{e^i\}\mapsto\{e'^i=A^i_je^j\}$ of the frames in $V$ is such
that $A^{-1}(x)=I(b)$ for all $x\in\base$ and some $b\in\field^s$.

\item
 $U \colon G\ni g_b\mapsto U(b)$, where $U(b)$ is an operator on the space of
state vectors, and the changes $(u,\{e^i\})\mapsto(u',\{e'^i\})$ of the
reference frames entail~\eref{2.8-2} with $U(b)$ for $U$.
    \end{enumerate}

	Under the above hypotheses equation~\eref{6.24} transforms into
    \begin{equation}    \label{6.25}
U(b)\circ\varphi_{u,i}(r)\circ U^{-1}(b)
=
\det\Bigl[ \frac{\pd (H_b(r))^i}{\pd r^j} \Bigr]
I_i^j(b) \varphi_{u,j}(H_b(r))
    \end{equation}
which can be called global Heisenberg relation in the particular situation. The
next step is to differentiate this equation with respect to $b^\omega$,
$\omega=1,\dots,s$, and then to put  $b=b_0$ in the result. In this way we
obtain the following (local) Heisenberg relation
    \begin{equation}    \label{6.26}
[U_\omega,\varphi_{u,i}(r)]_{\_}
=
  \Delta_\omega(r)\varphi_{u,i}(r)
+ I_{i\omega}^j \varphi_{u,j}(r)
+ (h_{\omega(r)})^k \frac{\pd \varphi_{u,i}(r)}{\pd r^k} ,
    \end{equation}
where
    \begin{subequations}    \label{6.27}
    \begin{align}    \label{6.27a}
U_\omega
& := \frac{\pd U(b)}{\pd b^\omega} \Big|_{b=b_0}
\\		    \label{6.27b}
\Delta_\omega(r)
& :=
\left.
\frac{\pd \det\Bigl[ \frac{\pd (H_b(r))^j}{\pd r^j} \Bigr] } {\pd b^\omega}
\right|_{b=b_0}
\in  \field[R]^{\dim \base}
\\		    \label{6.27c}
I_{i\omega}^j
& :=
\frac{\pd I_i^j(b)}{\pd b^\omega} \Big|_{b=b_0} \in\field
\\		    \label{6.27d}
h_\omega
& :=
\frac{\pd H_b}{\pd b^\omega} \Big|_{b=b_0}
 \colon \field[R]^{\dim \base}\to \field[R]^{\dim \base} .
    \end{align}
    \end{subequations}

	In particular, if $H_b$ is linear and non-homogeneous, \ie
 $H_b(r)=H(b)\cdot r + a(b)$ for some $H(b)\in\GL(\dim\base,\field[R])$ and
$a(b)\in\field^{\dim\base}$ with $H(b_0)=\openone$ and $a(b_0)=\Mat{0}$, then
($\Tr$ means trace of a matrix or operator)
    \begin{equation}    \label{6.28}
\Delta_\omega(r)
= \frac{\pd \det(H(b))}{\pd b_\omega} \Big|_{b=b_0}
= \frac{\pd \Tr(H(b))}{\pd b_\omega} \Big|_{b=b_0}
\qquad
h_\omega(\,\cdot\,)
= \frac{\pd H(b)}{\pd b_\omega} \Big|_{b=b_0} \cdot  (\,\cdot\,)
+ \frac{\pd a(b)}{\pd b_\omega} \Big|_{b=b_0}
    \end{equation}
as $\frac{\pd \det B}{\pd b_i^j}\big|_{B=\openone}=\delta_j^i$ for any square
matrix $B=[b_i^j]$. In this setting the Heisenberg relations corresponding to
Poincar\'e transformations (see subsection~\ref{Subsect6.1}) are described by
$b\mapsto(\Lambda^{\mu\nu},a^\lambda)$, $H(b)\mapsto\Lambda$, $a(b)\mapsto a$
and $I(b)\mapsto I(\Lambda)$, so that
 $U_\omega\mapsto(S_{\mu\nu,T_\lambda})$,
 $\Delta_\omega(r)\equiv 0$,
$I_{i\omega}^j\mapsto(I_{i \mu\nu}^j,0)$ and
\(
(h_\omega(r))^k\frac{\pd }{\pd r^k}
\mapsto
r_\mu\frac{\pd }{\pd r^\nu} - r_\nu\frac{\pd }{\pd r^\mu} .
\)

	The case of internal transformations, considered in the previous
subsection, corresponds to $H_b=\id_{\field[R]^{\dim\base}}$ and, consequently,
in it $\Delta_\omega(r)\equiv 0$ and $h_\omega=0$.

\section{Fibre bundle approach}
\label{Subsect6.4}

	Suppose a physical field is described as a section
$\varphi \colon \base\to E$ of a vector bundle $(E,\pi,\base)$. Here $\base$ is
a real differentiable (4\ndash)manifold (of class at least $C^1$), serving as a
spacetime model, $E$ is the bundle space and $\pi \colon \base\to E$ is the
projection; the fibres $\pi^{-1}(x)$, $x\in\base$, are isomorphic vector
spaces.~%

	Let $(U,u)$ be a chart of $\base$ and $\{e^i\}$ be a (vector) frame in
the bundle with domain containing $U$, \ie
 $e^i \colon x\mapsto e^i(x)\in\pi^{-1}(x)$ with $x$ in the domain of $\{e^i\}$
and $\{e^i(x)\}$  being a basis in $\pi^{-1}(x)$. Below we assume
$x\in U\subseteq \base$. Thus, we have
    \begin{equation}    \label{3.1}
\varphi \colon \base\ni x\mapsto \varphi(x)
=
\varphi_i(x) e^i(x)
=
\varphi_{u,i}(\Mat{x}) e^i(u^{-1}(\Mat{x})) ,
    \end{equation}
where
    \begin{equation}    \label{3.2}
\Mat{x}:=u(x) \qquad
\varphi_{u,i}:=\varphi_i\circ u^{-1}
    \end{equation}
and $\varphi_i(x)$ are the components of the vector
$\varphi(x)\in\pi^{-1}(x)$ relative to the basis $\{e^i(x)\}$ in $\pi^{-1}(x)$.

	The origin of the Heisenberg relations on the background of fibre
bundle setting is in the equivalent equations
	\begin{align}	\label{6.29}
U\circ\varphi_i(x)\circ U^{-1}
&= (A^{-1})_i^j(x) \varphi_j(x)
\\			\tag{\ref{6.29}$'$}
			\label{6.29'}
U\circ\varphi_{u,i}(\Mat{x})\circ U^{-1}
&= (A^{-1})_i^j(x) \varphi_{u,j}(\Mat{x}) .
	\end{align}

	Similarly to subsection~\ref{Subsect6.3}, consider a Lie group $G$, its
representations $I$ and $U$ and reference frames with the following properties:
    \begin{enumerate}
\item
$I \colon G\ni g_b\mapsto I(b)\in\GL(\dim V,\field)$
and the changes $\{e^i\}\mapsto\{e'^i=A^i_je^j\}$ of the frames in $V$ are such
that $A^{-1}(x)=I(b)$ for all $x\in\base$ and some $b\in\field^s$.

\item
 $U \colon g\ni g_b\mapsto U(b)$, where $U(b)$ is an operator on the space of
state vectors, and the changes $(u,\{e^i\})\mapsto(u',\{e'^i\})$ of the
reference frames entail~\eref{2.8-2} with $U(b)$ for $U$.
    \end{enumerate}

    \begin{Rem}    \label{Rem6.1}
One can consider also simultaneous coordinate changes
$u\mapsto u'=H_b\circ u$ induced by a representation
\(
H \colon G\ni g_b\mapsto H_b
   \colon \field[R]^{\dim\base}\to \field[R]^{\dim\base} ,
\)
as in subsection~\ref{Subsect6.3}. However such a supposition does not influence
our results as the basic equations~\eref{6.30} and~\eref{6.30'} below are
independent from it; in fact, equation~\eref{6.30} is coordinate-independent,
while~\eref{6.30'} is its version valid in any local chart $(U,u)$ as
$\varphi_{u}:=\varphi\circ u^{-1}$ and $\Mat{x}:=u(x)$.

    \end{Rem}

Thus equations~\eref{6.29} and~\eref{6.29'} transform into (cf.~\eref{6.25})
	\begin{align}	\label{6.30}
U(b)\circ\varphi_i(x)\circ U^{-1}(b)
&= I_i^j(b) \varphi_j(x)
\\			\tag{\ref{6.30}$'$}
			\label{6.30'}
U(b)\circ\varphi_{u,i}(\Mat{x})\circ U^{-1}(b)
&= I_i^j(b) \varphi_{u,j}(\Mat{x}) .
	\end{align}
Differentiating~\eref{6.30} with respect to $b^\omega$ and then putting
$b=b_0$, we derive the following Heisenberg relation
	\begin{align}	\label{6.31}
[U_\omega,\varphi_i(x)]_{\_}
&= I_{i\omega}^j \varphi_j(x)
\\\intertext{or its equivalent version (cf.~\eref{6.26})}
			\tag{\ref{6.31}$'$}
			\label{6.31'}
[U_\omega,\varphi_{u,i}(\Mat{x})]_{\_}
&= I_{i\omega}^j \varphi_{u,j}(\Mat{x}) ,
	\end{align}
where
    \begin{subequations}    \label{6.32}
    \begin{align}    \label{6.32a}
U_\omega
& := \frac{\pd U(b)}{\pd b^\omega} \Big|_{b=b_0}
\\		    \label{6.32b}
I_{i\omega}^j
& :=
\frac{\pd I_i^j(b)}{\pd b^\omega} \Big|_{b=b_0} .
    \end{align}
    \end{subequations}
We can rewire the Heisenberg relations obtained as
    \begin{equation}    \label{6.33}
[U_\omega,\varphi]_{\_} = I_{i\omega}^{j} \varphi_j e^i .
    \end{equation}
One can prove that the r.h.s.\ of this equation is independent of the
particular frame $\{e^i\}$ in which it is represented.

	The case of Poncar\'e transformations is described by the replacements
$b\mapsto(\Lambda^{\mu\nu},a^\lambda)$, $U_\omega\mapsto(S_{\mu\nu},T_\lambda)$
and $I_{i\omega}^{j}\mapsto(I_{i\mu\nu}^j,0)$ and, consequently, the
equations~\eref{6.30} and~\eref{6.30'} now read
	\begin{align}	\label{6.33-1}
U(\Lambda,a)\circ\varphi_i(x)\circ U^{-1}(\Lambda,a)
&= I_i^j(\Lambda,a) \varphi_j(x)
\\			\tag{\ref{6.33-1}$'$}
			\label{6.30-1'}
U(\Lambda,a)\circ\varphi_{u,i}(\Mat{x})\circ U^{-1}(\Lambda,a)
&= I_i^j(\Lambda,a) \varphi_{u,j}(\Mat{x}) .
	\end{align}
Hence, for instance, the Heisenberg relations~\eref{6.31} now takes the form
(cf.~\eref{6.7})
    \begin{subequations}    \label{6.34}
    \begin{align}    \label{6.34a}
[T_{\mu},\varphi_i(x)]_{\_} &= 0
\\		    \label{6.34b}
[S_{\mu\nu},\varphi_i(x)]_{\_} &= I_{i\mu\nu}^{j} \varphi_j(x) .
    \end{align}
    \end{subequations}
Respectively, the correspondences~\eref{6.9} transform these equations into
    \begin{subequations}    \label{6.35}
    \begin{align}    \label{6.35a}
[P_{\mu},\varphi_i(x)]_{\_} &= 0
\\		    \label{6.35b}
[M_{\mu\nu},\varphi_i(x)]_{\_} &= I_{i\mu\nu}^{j} \varphi_j(x)
    \end{align}
    \end{subequations}
which now replace~\eref{6.10}.

	Since equation~\eref{6.10a} (and partially equation~\eref{6.10b}) is
(are) the corner stone for the particle interpretation of quantum field
theory~\cite{Bogolyubov&Shirkov,Bjorken&Drell,bp-MP-book}, the
equation~\eref{6.35a} (and partially equation~\eref{6.35b}) is (are) physically
unacceptable if one wants to retain the particle interpretation in the fibre
bundle approach to the theory. For this reason, it seems that the
correspondences~\eref{6.9} should \emph{not} be accepted in the fibre bundle
approach to quantum field theory, in which~\eref{6.7} transform
into~\eref{6.34}. However, for retaining the particle interpretation one can
impose~\eref{6.10} as subsidiary restrictions on the theory in the fibre bundle
approach. It is almost evident that this is possible if the frames used are
connected by linear homogeneous transformations with spacetime constant
matrices, $A(x)=\const$ or $\pd_\mu A(x)=0$. Consequently, if one wants to
retain the particle interpretation of the theory, one should suppose the
validity of~\eref{6.10} in some frame and, then, it will hold in the whole
class of frames obtained from one other by transformations with spacetime
independent matrices.

	Since the general setting investigated above is independent of any
(local) coordinates, it describes also the fibre bundle version of the case of
internal transformations considered in section~\ref{Subsect6.2}. This
explains why equations like~\eref{6.15} and~\eref{6.31'} are identical but the
meaning of the quantities $\varphi_{u,i}$ and $I_{i\omega}^{j}$ in them is
different.~%
\footnote{~%
Note, now $I(b)$ is the matrix defining transformations of frames in the bundle
space, while in~\eref{6.19} it serves a similar role for frames in the vector
space $V$.%
}
In particular, in the case of phase transformations
    \begin{equation}    \label{6.36}
U(b) = \e^{\frac{1}{\iu e} b Q_1}
\quad
I(b) = \openone \e^{-\frac{q}{\iu e} b}
\qquad
b\in\field[R]
    \end{equation}
the Heisenberg relations~\eref{6.31} reduce to
    \begin{equation}    \label{6.37}
[Q_1,\varphi_i(x)]_{\_} = -q \varphi_i(x),
    \end{equation}
which is identical with~\eref{6.21}, but now $\varphi_i$ are the components of
the section $\varphi$ in $\{e^i\}$. The invariant form of the last relations is
    \begin{equation}    \label{6.38}
[Q_1,\varphi]_{\_} = -q \varphi
    \end{equation}
which is also a consequence from~\eref{6.33} and~\eref{6.20}.



\section {Conclusion}
\label{Conclusion}

	In this paper we have shown how the Heisenberg equations arise in the
general case and in particular situations. They are from pure
geometrical origin and one should be careful when applying them to the
Lagrangian formalism in which they are subsidiary conditions, like the Lorentz
gauge in the electrodynamics. In the general case they need not to be consistent
with the Lagrangian formalism and their validity should carefully be checked.
For instance, if one starts with field operators in the Lagrangian formalism of
free fields and adds to it the Heisenberg relations~\eref{6.10a} concerning the
momentum operator, then the arising scheme is not consistent as in it start to
appear distributions, like the Dirac delta function. This conclusion leads to
the consideration of the quantum fields as operator-valued distribution in the
Lagrangian formalism even for free fields. In the last case, the Heisenberg
relations concerning the momentum operator are consistent with the Lagrangian
formalism. Besides, they play an important role in the particle interpretation
of the so-arising theory.


\section*{Acknowledgments}

	This work was partially supported by the National Science Fund of
Bulgaria under Grant No.~F~1515/2005.


\addcontentsline{toc}{section}{References}
\bibliography{bozhopub,bozhoref}
\bibliographystyle{unsrt}
\addcontentsline{toc}{subsubsection}{This article ends at page}



\end{document}